\documentclass[12pt]{article}
\usepackage{amssymb}
\usepackage{amsfonts}
\usepackage{amsmath}
\usepackage{natbib}
\usepackage{graphicx}

\newtheorem{algorithm}{Algorithm}

\date{February 19, 2013}
\begin{document}

\title{Bayesian Inference for Logistic Regression Models using Sequential
Posterior Simulation}
\author{John Geweke\thanks{%
University of Technology Sydney (Australia), Erasmus University (The
Netherlands) and University of Colorado (US). Support from Australian
Research Council grant 130103356 is gratefully acknowledged. Corresponding
author: John.Geweke@uts.edu.au}, Garland Durham\ and Huaxin Xu\thanks{%
University of Technology Sydney }}
\maketitle

\begin{abstract}
The logistic specification has been used extensively in non-Bayesian
statistics to model the dependence of discrete outcomes on the values of
specified covariates. Because the likelihood function is globally weakly
concave estimation by maximum likelihood is generally straightforward even
in commonly arising applications with scores or hundreds of parameters. In
contrast Bayesian inference has proven awkward, requiring normal
approximations to the likelihood or specialized adaptations of existing
Markov chain Monte Carlo and data augmentation methods. This paper
approaches Bayesian inference in logistic models using recently developed
generic sequential posterior simulaton (SPS) methods that require little
more than the ability to evaluate the likelihood function. Compared with
existing alternatives SPS is much simpler, and provides numerical standard
errors and accurate approximations of marginal likelihoods as by-products.
The SPS algorithm for Bayesian inference is amenable to massively parallel
implementation, and when implemented using graphical processing units it is
more efficient than existing alternatives. The paper demonstrates these
points by means of several examples.
\end{abstract}

\bigskip \pagebreak

\section{Introduction\label{sec:introduction}}

The multinomial logistic regression model, hereafter \textquotedblleft logit
model,\textquotedblright\ is one of the most widely used models in applied
statistics. It provides one of the simplest and most straightforward links
between the probabilities of discrete outcomes and covariates. More
generally, it provides a workable probability distribution for discrete
events, whether directly observed or not, as a function of covariates. In
the latter, more general, setting it is a key component of conditional
mixture models including the mixture of experts models introduced by Jacobs
et al. (1991) and studied by Jiang and Tanner (1999).

The logit model likelihood function is unimodal and globally concave, and
consequently estimation by maximum likelihood is practical and reliable even
in models with many outcomes and many covariates. However, it has proven
less tractable in a Bayesian context, where effective posterior simulation
has been a challenge. Because it also arises frequently in more complex
contexts like mixture models, this is a significant impediment to the
penetration of posterior simulation methods. Indeed, the multinomial probit
model has proven more amenable to posterior simulation methods (Albert and
Chib, 1993; Geweke et al., 1994) and has sometimes been used in lieu of the
logit model in conditional mixture models (Geweke and Keane, 2007). Thus
there is a need for simple and reliable posterior simulation methods for
logit models.

State-of-the-art approaches to posterior simulation for logit models use
combinations of likelihood function approximation and data augmentation in
the context of Markov chain Monte Carlo (MCMC) algorithms: see Holmes and
Held (2006), Fr\"{u}hwirth and Fr\"{u}hwirth-Schnatter (2007), Scott (2011),
Gramacy and Polson (2012) and Polson et al. (2012). The last paper uses a
novel representation of latent variables based on Polya-Gamma distributions
that can be applied in logit and related models, and uses this
representation to develop posterior simulators that are reliable and
substantially dominate alternatives with respect to computational
efficiency. Going forward, we refer to the method of Polson et al. (2012) as
the PSW algorithm.

This paper implements a sequential posterior simulator (SPS) using ideas
developed in Durham and Geweke (2012). Unlike MCMC this algorithm is
especially well-suited to massively parallel computation using graphical
processing units (GPUs). The algorithm is highly generic; that is, the
coding effort required to adapt it to a specific model is typically minimal.
In particular, the algorithm is far simpler to implement for the logit
models considered here than the existing MCMC algorithms mentioned in the
previous paragraph. When implemented on GPUs the computational efficiency of
SPS compares favorably with existing MCMC methods, but even on ubiquitous
quadcore machines it is still competitive, if slower. Moreover, SPS yields
an accurate approximation of log marginal likelihood, as well as reliable
and systematic indications of the accuracy of posterior moment
approximations, which existing MCMC methods do not.

Section \ref{sec:SMC} of the paper describes the SPS algorithm and its
implementation on GPUs, with emphasis on the specifics of the logit model.
Section \ref{sec:Models} provides the background for the examples taken up
subsequently: specifics of the models and prior distributions, the data
sets, and the various hardware and software environments used. Section \ref%
{sec:SMC_performance} documents some of the features of models and data sets
that govern computation time using SPS. Section \ref{sec:comparison} studies
several applications of the logit model using data sets typical of applied
work in biostatistics and the social sciences. Section 6 concludes with some
more general observations. A quick first reading of the paper might skim
Section 2 and skip Section 4.

\section{Sequential Monte Carlo algorithms for Bayesian inference\label%
{sec:SMC}}

Sequential posterior simulation (SPS) grows out of sequential Monte Carlo
(SMC) methods developed over the past twenty years for state space models,
in particular particle filters. Contributions leading up to the development
here include Baker (1985, 1987), Gordon et al. (1993), Kong et al. (1994),
Liu and Chen (1995, 1998), Chopin (2002, 2004), Del Moral et al. (2006),
Andreiu et al. (2010), Chopin and Jacob (2010), and Del Moral et al. (2011).
Despite its name, SPS is amenable to massively parallel implementation. It
is nearly ideally suited to graphical processing units, which provide
massively parallel desktop computing at a cost of well under one dollar (US)
per processing core. For further background on these details see Durham and
Geweke (2012).

\subsection{Conditions\label{subsec:Conditions}}

The multinomial logit model assigns probabilities to random variables $%
Y_{t}\in \left\{ 1,2,\ldots ,C\right\} $ as functions of observed covariates 
$x_{t}$ and a parameter vector $\theta $. In the simplest setup $\theta
^{\prime }=\left( \theta _{1}^{\prime },\ldots ,\theta _{C}^{\prime }\right) 
$ and 
\begin{equation}
P\left( Y_{t}=c\mid x_{t},\theta \right) =\frac{\exp \left( \theta
_{c}^{\prime }x_{t}\right) }{\sum_{i=1}^{C}\exp \left( \theta _{i}^{\prime
}x_{t}\right) }\text{ }\left( c=1,\ldots ,C;t=1,\ldots ,T\right) \text{.%
\label{plogit}}
\end{equation}%
There is typically a normalization $\theta _{c}=0$ for some $c\in \left\{
1,2,\ldots ,C\right\} $, and there could be further restrictions on $\theta $%
, but these details are not important to the main points of this section.
For the properties of the SPS algorithm discussed subsequently what matters
is that the likelihood function implied by (\ref{plogit}) is (a) bounded and
(b) easy to evaluate to machine accuracy. The SPS\ algorithm also requires
(c) a proper prior distribution from which one can simulate $\theta $. We
will confine ourselves to the approximation of posterior moments of the form 
\begin{equation*}
\overline{g}=\mathrm{E}\left[ g\left( \theta \right) \mid y_{1:T},x_{1:T}%
\right] 
\end{equation*}%
for which (d) $\mathrm{E}\left[ g\left( \theta \right) ^{2+\delta }\right] $
(some $\delta >0$) exists under the prior distribution. For example, if the
prior distribution is Gaussian then condition (c) is satisfied and if the
function of interest is a log odds-ratio evaluated at a particular value of
the covariate vector, then condition (d)\ is also met.

For many posterior simulation algorithms yielding a sequence $\left\{ \theta
_{n}\right\} $, it is known that under these conditions the mean from a
sample of size $N$ from the simulator, $\overline{g}^{N}=N^{-1}%
\sum_{n=1}^{N}g\left( \theta _{n}\right) $, satisfies a central limit theorem%
\begin{equation}
N^{1/2}\left( \overline{g}^{N}-\overline{g}\right) \overset{d}{%
\longrightarrow }N\left( 0,v\right) \text{.\label{basic_CLT}}
\end{equation}%
For example, this is the case in many, arguably most, implementations of the
Metropolis-Hastings random walk algorithm. It is also true for the SPS
algorithm detailed in the next section.

Prudent use of posterior simulation requires that it be possible to compute
a simulation-consistent approximation of $v$ in (\ref{basic_CLT}), $v^{N}%
\overset{a.s.}{\longrightarrow }v$. This has proved difficult in the case of
MCMC (Flegal and Jones, 2010) and it appears the problem has been ignored in
the SMC literature. But, as suggested by Durham and Geweke (2012), one can
always work around this difficulty by undertaking $J$ independent
replications of the algorithm. Given posterior draws $g_{jn}=g\left( \theta
_{jn}\right) $ $\left( j=1,\ldots ,J;\text{ }n=1,\ldots ,N\right) $, group
means are given by 
\begin{equation}
\overline{g}_{j}^{N}=N^{-1}\sum_{n=1}^{N}g_{jn}\text{ \ }\left( j=1,\ldots
,J\right) \text{;\label{partial_g}}
\end{equation}%
and from (\ref{basic_CLT}) satisfy 
\begin{equation}
N^{1/2}\left( \overline{g}_{j}^{N}-\overline{g}\right) \overset{d}{%
\longrightarrow }N\left( 0,v\right) \text{ \ }\left( j=1,\ldots ,J\right) 
\text{.\label{partial_g_CLT}}
\end{equation}%
For approximation of the posterior moment we can then use the grand mean 
\begin{equation}
\overline{g}^{\left( J,N\right) }=J^{-1}\sum_{j=1}^{J}\overline{g}_{j}^{N}%
\text{,\label{whole_g}}
\end{equation}%
which suggests using the natural approximation of $v$ in (\ref{partial_g_CLT}%
), 
\begin{equation}
\widehat{v}^{\left( J,N\right) }\left( g\right) =\left[ N/\left( J-1\right) %
\right] \sum_{j=1}^{J}\left( \overline{g}_{j}^{N}-\overline{g}^{\left(
J,N\right) }\right) ^{2}\text{\label{v-hat}}
\end{equation}%
to approximate $v$ in (\ref{basic_CLT}). 

We will define the numerical standard error of $\overline{g}^{\left(
J,N\right) }$%
\begin{equation}
\mathrm{NSE}\left( \overline{g}^{\left( J,N\right) }\right) =\left[ J^{-1}%
\widehat{v}^{\left( J,N\right) }\left( g\right) \right] ^{1/2}\text{.\label%
{NSE_def}}
\end{equation}%
As $N\rightarrow \infty $ 
\begin{equation}
\left( J-1\right) \widehat{v}^{\left( J,N\right) }\left( g\right) /v\overset{%
d}{\rightarrow }\chi ^{2}\left( J-1\right) \text{\label{v-hat_limit}}
\end{equation}%
and%
\begin{equation}
\frac{\overline{g}^{\left( J,N\right) }-\overline{g}}{\mathrm{NSE}\left( 
\overline{g}^{\left( J,N\right) }\right) }\overset{d}{\rightarrow }t\left(
J-1\right) \text{.\label{gbar_tstat}}
\end{equation}%
The NSE provides a measure of the variability of the moment approximation (%
\ref{whole_g}) across replications of the algorithm with fixed data.

The relative numerical efficiency (RNE; Geweke 1989), which approximates the
population moment ratio $\mathrm{var}\left( g\left( \theta \right) \mid
y_{1:T}\right) /v$, can be obtained in a similar manner, 
\begin{equation}
\mathrm{RNE}\left( \overline{g}^{\left( J,N\right) }\right) =\left(
JN\right) ^{-1}\sum_{j=1}^{J}\sum_{n=1}^{N}\left( g_{jn}-\overline{g}%
^{\left( J,N\right) }\right) ^{2}/\widehat{v}^{\left( J,N\right) }\left(
g\right) \text{.\label{RNE_def}}
\end{equation}%
RNE close to one indicates that there is little dependence amongst the
particles, and that the moment approximations (3) and (5) approach the
efficiency of the ideal, an iid sample from the posterior. RNE less than one
indicates dependency. In this case, the moment approximations (3) and (5)
are less precise than one would obtain with a hypothetical iid sample.

This is all rather awkward for MCMC, requiring as it does $J$ repetitions of
the algorithm complete with burn-in; we use it in Sections 3 and 4 to assess
the accuracy of posterior moment approximations of the Polson et al. (2012)
procedure, in the absence of a better alternative. However, the procedure is
natural in the context of the SPS algorithm, requires no additional
computations, and makes efficient use of massively parallel computing
environments (Durham and Geweke, 2012). 

Going forward in this section, $p\left( \theta \right) $ denotes the prior
density of $\theta $. The vectors $y_{1},\ldots ,y_{T}$ denote the data and $%
y_{1:t}=\left\{ y_{1},\ldots ,y_{t}\right\} $. The notation suppresses
conditioning on the covariates $x_{t}$ and treats all distributions as
continuous to avoid notational clutter. Thus, for example, the likelihood
function is%
\begin{equation*}
p\left( y_{1:T}\mid \theta \right) =\prod\limits_{t=1}^{T}p\left( y_{t}\mid
y_{1:t-1},\theta \right) \text{.}
\end{equation*}

\subsection{\label{subsec:Nonadaptive}Non-adaptive SPS algorithms}

We begin with a mild generalization of the SMC algorithm of Chopin (2004).
The algorithm generates and modifies different values of the parameter
vector $\theta $, known as particles and denoted $\theta _{jn}$, with
superscripts used for further specificity at various points in the
algorithm. The subscripts refer to the $J$ groups of $N$ values each
described in the previous section. To make the notation compact, let $%
\mathcal{J}\!=\left\{ 1,\ldots ,J\right\} \,\ $and $\mathcal{N=}\left\{
1,\ldots ,N\right\} $. The algorithm is an implementation of Bayesian
learning, providing simulations from $\theta \mid y_{1:t}$ for $t=1,2,\ldots
,T$. It processes observations, in order and in successive batches, each
batch constituting a cycle of the algorithm.

The global structure of the algorithm is therefore iterative, proceeding
through the sample. But it operates on many particles in exactly the same
way at almost every stage, and it is this feature of the algorithm that
makes it amenable to massively parallel implementations. On conventional
quadcore machines and samples of typical size one might set up the algorithm
with $J=10$ groups of $1000$ particles each, and using GPUs $J=40$ groups of 
$2500$ particles each. (The numbers are just illustrations, to fix ideas.)

\begin{algorithm}
\label{alg:general_nonadaptive}(Nonadaptive) \ Let $t_{0},\ldots ,t_{L}$ be
fixed integers with $0=t_{0}<t_{1}<\ldots <t_{L}=T$; these define the cycles
of the algorithm. Let $\lambda _{1},\ldots ,\lambda _{L}$ be fixed vectors
that parameterize transition distributions as indicated below.
\end{algorithm}

\begin{enumerate}
\item Initialize $\ell =0$ and let $\theta _{jn}^{\left( \ell \right) }%
\overset{iid}{\thicksim }p\left( \theta \right) \quad \left( j\in \mathcal{J}%
\!,\,n\in \mathcal{N}\right) $

\item For $\ell =1,\dots,L$

\begin{enumerate}
\item Correction $\left( C\right) $ phase, for all $j\in \mathcal{J}$ and $%
\,n\in \mathcal{N}$:

\begin{enumerate}
\item $w_{jn}\left( t_{\ell -1}\right) =1$

\item For $s=t_{\ell -1}+1,\dots ,t_{\ell }$%
\begin{equation}
w_{jn}\left( s\right) =w_{jn}\left( s-1\right) \cdot p\left( y_{s}\mid
y_{1:s-1},\theta _{jn}^{\left( \ell -1\right) }\right)
\label{C_phase_compute}
\end{equation}

\item $w_{jn}^{\left( \ell -1\right) }:=w_{jn}\left( t_{\ell }\right) $
\end{enumerate}

\item Selection $\left( S\right) $ phase, applied independently to each
group $j\in \mathcal{J}$: Using multinomial or residual sampling based on $%
\left\{ w_{jn}^{\left( \ell -1\right) }\,\left( n\in \mathcal{N}\right)
\right\} $, select 
\begin{equation*}
\left\{ \theta _{jn}^{\left( \ell ,0\right) }\,\left( n\in \mathcal{N}%
\right) \right\} \text{ from }\left\{ \theta _{jn}^{\left( \ell -1\right)
}\,\left( n\in \mathcal{N}\right) \right\}
\end{equation*}

\item Mutation $\left( M\right) $ phase, applied independently across $j\in 
\mathcal{J}\!,\,n\in \mathcal{N}$: 
\begin{equation}
\theta _{jn}^{\left( \ell \right) }\thicksim p\left( \theta \mid
y_{1:t_{\ell }},\theta _{jn}^{\left( \ell ,0\right) },\lambda _{\ell
}\right) \text{\label{Mphase_pdf}}
\end{equation}%
where the drawings are independent and the p.d.f.~(\ref{Mphase_pdf})
satisfies the invariance condition 
\begin{equation}
\int_{\Theta }p\left( \theta \mid y_{1:t_{\ell }},\theta ^{\ast },\lambda
_{\ell }\right) p\left( \theta ^{\ast }\mid y_{1:t_{\ell }}\right) d\nu
(\theta ^{\ast })=p\left( \theta \mid y_{1:t_{\ell }}\right)
\label{M_invariant}
\end{equation}
\end{enumerate}

\item $\theta _{jn}:=\theta _{jn}^{\left( L\right) }\quad \left( j\in 
\mathcal{J}\!,\,n\in \mathcal{N}\right) $
\end{enumerate}

The algorithm is nonadaptive because $t_{0},\ldots ,t_{L}$ and $\lambda
_{1},\ldots ,\lambda _{L}$ are fixed before the algorithm starts. Going
forward it will be convenient to denote the cycle indices by $\mathcal{L}%
=\left\{ 1,\ldots ,L\right\} $. At the conclusion of the algorithm, the
simulation approximation of a generic posterior moment is (\ref{whole_g}).

The only communication between particles is in the $S$ phase. In the $C$ and 
$M$ phases exactly the same computations are made for each particle, with no
communication. This situation is ideal for GPUs, as detailed in Durham and
Geweke (2012). In the $S$ phase there is communication between particles
within, but not across, the $J$ groups. This keeps the particles in the $J$
groups independent. Typically the fraction of computation time devoted to
the $S$ phase is minute.

For each group, $j\in \mathcal{J}\!$, the four regularity conditions in the
previous section imply the assumptions of Chopin (2004), Theorem 1 (for
multinomial resampling) and Theorem 2 (for residual resampling). Therefore a
central limit theorem (\ref{basic_CLT}) applies. Chopin provides population
expressions for $v$ in terms of various unknown moments but neither that
paper, nor to our knowledge any other paper, provides a way to approximate $%
v $.

The approach described in the previous section solves this problem. \ Notice
that dependence amongst the particles arises solely from the $S$ phase, in
which resampling is applied independently to each group $j\in \mathcal{J}$:
Therefore the $J$ partial means $\overline{g}_{j}^{N}$ are mutually
independent for all $N$. The procedures for approximating $v$, numerical
standard errors, and a large-$N$ theory of numerical accuracy laid out in (%
\ref{partial_g}) - (\ref{gbar_tstat}) therefore apply.

\subsection{Adaptive SPS algorithms\label{subsec:Adaptive}}

In Algorithm \ref{alg:general_nonadaptive} neither the cycles, defined by $%
t_{1},\ldots ,t_{L-1}$, nor the hyperparameters $\lambda _{\ell }$ of the
transition processes (\ref{Mphase_pdf}) depend on the particles $\left\{
\theta _{jn}\right\} $. With respect to the random processes that generate
these particles, these hyperparameters are fixed: in econometric
terminology, they are predetermined with respect to $\left\{ \theta
_{jn}\right\} $. As a practical matter, however, one must use the knowledge
of the posterior distribution inherent in the particles to choose the
transition from the $C$ phase to the $S$ phase, and to design an effective
transition distribution in the $M$ phase. Without this feedback it is
impossible to obtain an approximation $\overline{g}^{\left( J,N\right) }$ of 
$\overline{g}$ with acceptably small NSE; indeed, in all but the simplest
models and smallest data sets $\overline{g}^{\left( J,N\right) }$ will
otherwise be pure noise, for all intents and purposes.

The following procedure illustrates how the particles themselves can be used
to choose the cycles defined by $t_{1},\ldots ,t_{L-1}$ and the
hyperparameters $\lambda _{\ell }$ of the transition processes. It is a
minor modification of a procedure first described in Durham and Geweke
(2012), that has proved effective in a number of models. It is also
effective in the logit model. The algorithm requires that the user choose
the number of groups, $J$, and the number of particles in each group, $N$.

\begin{algorithm}
\label{alg:specific_adaptive} (Adaptive)
\end{algorithm}

\begin{enumerate}
\item \label{Alg_adaptive_start}Determine the value of $t_{\ell }$ in the $C$
phase of cycle $\ell \,\left( \ell \in \mathcal{L}\right) $ as follows.

\begin{enumerate}
\item At each step $s$ compute the effective sample size 
\begin{equation}
\mathit{ESS}\left( s\right) =\frac{\left[ \sum_{j=1}^{J}\sum_{n=1}^{N}w_{jn}%
\left( s\right) \right] ^{2}}{\sum_{j=1}^{J}\sum_{n=1}^{N}w_{jn}\left(
s\right) ^{2}}\text{\label{ESS_rule}}
\end{equation}%
immediately after computing \eqref{C_phase_compute}.

\item \label{ESS_step}If $\mathit{ESS}\left( s\right) /\left( J\cdot
N\right) <0.5$ or if $s=T$ set $t_{\ell }=s$ and proceed to the $S$ phase.
Otherwise increment $s$ and recompute (\ref{ESS_rule}).
\end{enumerate}

\item The transition density (\ref{Mphase_pdf}) in the $M$ phase of each
cycle $\ell $ is a Metropolis Gaussian random walk, executed in steps $%
r=1,2,\ldots $ .

\begin{enumerate}
\item Initializiations:

\begin{enumerate}
\item $r=1$.

\item If $\ell =1$ then the step size scaling factor $h_{11}=0.5$.
\end{enumerate}

\item Set $\mathrm{RNE}$ termination criteria:

\begin{enumerate}
\item \label{step_2b}If $s<T$, $K=0.35$

\item \label{Step_2blast}If $s=T$, $K=0.9$
\end{enumerate}

\item \label{Alg_adaptive_M}Execute the next Metropolis Gaussian random walk
step.

\begin{enumerate}
\item \label{failstep}Compute the sample variance $V_{\ell r}$ of the
particles 
\begin{equation*}
\theta _{jn}^{\left( \ell ,r-1\right) }\,\,\left( j=1,\ldots ,J;n=1,\ldots
,N\right) ,
\end{equation*}%
define $\Sigma _{\ell r}=h_{\ell r}\cdot \,V_{\ell r}$, and execute step $r$
using a random walk Gaussian proposal density with variance matrix $\Sigma
_{\ell r}$ to produce a new collection of particles $\theta _{jn}^{\left(
\ell ,r\right) }\,\left( j=1,\ldots ,J;\text{ }n=1,\ldots ,N\right) $. Let $%
\alpha _{\ell r}$ denote the Metropolis acceptance rate across all particles
in this step.

\item Set $h_{\ell ,r+1}=\min \left( h_{\ell r}+0.01,1.0\right) $ if $%
a_{\ell r}>0.25$ and $h_{\ell ,r+1}=\max \left( h_{\ell r}-0.01,0.1\right) $
otherwise.

\item Compute the RNE of the numerical approximation $\overline{g}^{\left(
J,N\right) }$ to a test function $g^{\ast }\left( \theta \right) $. If $%
\mathrm{RNE}<K$ then increment $r$ and return to step \ref{Alg_adaptive_M};
otherwise set $h_{\ell +1,1}=h_{\ell ,r+1}$, define $R_{\ell }=r$, and
return to step \ref{Alg_adaptive_start}.
\end{enumerate}

\item \label{Alg_adaptive_final}Set $\theta _{jn}^{(\ell )}=\theta
_{jn}^{(\ell ,r)}$. If $s<T$ \ then set $h_{\ell +1,1}=h_{\ell ,r+1}$ and
return to step \ref{Alg_adaptive_start}; otherwise set $\theta _{jn}=\theta
_{jn}^{(\ell )}$, define $L=\ell $, and terminate.
\end{enumerate}
\end{enumerate}

At each step of the algorithm particles are identically but not
independently distributed. As the number of particles in each group $%
N\rightarrow \infty $ the common distribution coincides with the posterior
distribution. As the number of Metropolis steps, $r$, in the $M$ phase
increases, dependence amongst particles decreases. The $M$ phase terminates
when the RNE criterion is satisfied, implying a specified degree of
independence for the particles at the end of each cycle. Larger values of $J$
provide better estimates of RNE, making this assessment more reliable. The
RNE criterion $K$ assures a specified degree of independence at the end of
each cycle. The assessment of numerical accuracy is based on the comparison
of different approximations in $J$ groups of particles, and larger values of 
$J$ make this assessment more reliable.

At the conclusion of the algorithm, the posterior moments of interest $%
E(g(\theta )|y_{1:T})$ are approximated, 
\begin{equation*}
\overline{g}^{\left( J,N\right) }=\left( JN\right)
^{-1}\sum_{j=1}^{J}\sum_{n=1}^{N}g\left( \theta _{jn}\right) \text{.}
\end{equation*}%
The asymptotic (in $N$) variance of the approximation is proportional to $%
\left( JN\right) ^{-1}$, and because $K=0.9$ in the last cycle $L$ the
factor of proportionality is approximately the posterior variance $\mathrm{%
var}\left( g\left( \theta \right) \mid y_{1:T}\right) $. As detailed in
Section \ref{subsec:Conditions}, the accuracy of the reported NSE is
proportional to $J^{-1/2}$.

The division of a given posterior sample size into a number of groups $J$
and particles within groups $N$ should be guided by the tradeoff implied by (%
\ref{v-hat_limit}) and the fact that values of $N$ sufficiently small will
give misleading representations of the posterior distribution. From (\ref%
{v-hat_limit}) notice that the ratio of squared $NSE$ from one simulation to
another has an asymptotic (in $N$) $F\left( J-1,J-1\right) $ distribution.
For $J=8$, the ratio of $NSE$ in two simulations will be less than 2 with
probability 0.95. A good benchmark for serviceable approximation of
posterior moments is $J=10$, $N=1000$. With implementation on GPUs much
larger values can be practical: Durham and Geweke (2012) use $J=64$ and $%
N=4096$ in an application that is computationally much more demanding than
the examples in this paper.

The statement of Algorithm 2 shows that it contains several algorithm design
parameters that are simply fixed. These fixed parameters have been set to
ensure that as the algorithm proceeds through the sample it maintains a
workable accuracy of approximation, and does so in a computationally
efficient manner.

Upon entering the $S$ phase, the effective sample size is less than half the
number of particles (except perhaps in the final cycle). After resampling,
the number of unique particles is roughly equal to the effective sample size
before resampling, but the ESS measure (\ref{ESS_rule}) is no longer valid
when applied to the new sample (since it does not account for dependence
between particles). In the $M$ phase, iterating the Metropolis step reduces
dependence between particles, and the RNE after each step provides a way of
assessing the effectiveness of the mixing that takes place. When RNE gets
close to one, further Metropolis steps are of little utility and a waste of
computing resources. Prior to the final cycle we have found it practical to
terminate the $M$ phase when RNE exceeds 0.35. In the final cycle, we have
found it useful to undertake additional $M$ steps in order to get higher RNE
(and lower NSE) when approximating the posterior moments of interest. We
suggest a criterion of $\text{RNE}>0.9$ for the final cycle. Since these
extra iterations occur only in the final step, the relative cost is low ---
indeed the cost is typically much lower than the alternative of increasing
the number of particles.

Performance of the algorithm is not very sensitive to changes in the ESS
criterion. Higher thresholds lead to more cycles but fewer iterations in the 
$M$ phase; lower values lead to fewer cycles but more time in the $M$ phase.
On the other hand, changing the RNE criteria has the effects one might
expect. Changing the RNE criterion in the final cycle affects the accuracy
of the posterior moment approximations. Changing the RNE criterion for the
other cycles does not affect accuracy of posterior moment approximations
undertaken at time $T$, but does affect the accuracy of the approximations
of log marginal likelihood and log predictive likelihoods. As detailed in
Durham and Geweke (2012) , Section 4, these approximations are based on the
particle representation of the intermediate posterior distributions $p\left(
\theta \mid y_{1:t}\right) $. Increasing 0.35 -- say, to 0.9 -- will
increase the accuracy of the approximation.  The effect is to reduce NSE for
log marginal likelihood by a factor of roughly $1-\left( 0.35/0.9\right)
^{1/2}\approxeq 1/3$, and total computing time can increase by 50\% or more.
We have found the constants suggested above to provide a good balance in the
various tradeoffs involved. But the software that supports the work in this
paper makes it convenient for the knowledgeable user to change any of the
\textquotedblleft hardwired" design parameters in Algorithm 2 if desired.

\subsection{The two-pass SPS algorithm\label{sec:Models_etc}}

Algorithm \ref{alg:specific_adaptive} is practical and reliable in a very
wide array of applications. This includes situations in which MCMC is
utterly ineffective, as illustrated in Durham and Geweke (2012) and Herbst
and Schorfheide (2012). However there is an important drawback: the
algorithm has no supporting central limit theorem.

The effectiveness of the algorithm is due in no small part to the fact that
the cycle definitions $\left\{ t_{\ell }\right\} $ and parameters $\lambda
_{\ell }$ of the $M$ phase transition distributions are based on the
particles themselves. This creates a structure of dependence amongst
particles that is extremely complicated. The degree of complication stemming
from the use of effective sample size in step \ref{ESS_step} can be managed:
see Del Moral et al. (2012). But the degree of complication introduced in
the $M$ phase, step \ref{Alg_adaptive_M}, is orders of magnitude larger.
This is not addressed by any of the relevant literature, and (in our view)
this problem is not likely to be resolved by attacking it directly anytime
in the foreseeable future.

Fortunately, the problem can be solved at the cost of roughly doubling the
computational burden in the following manner as proposed by Durham and
Geweke (2012).

\begin{algorithm}
\label{alg:two_pass}(Two pass)
\end{algorithm}

\begin{enumerate}
\item Execute the adaptive Algorithm \ref{alg:specific_adaptive}. Discard
the particles $\left\{ \theta _{jn}\right\} $. Retain the number of cycles $%
L $, values $t_{0},\ldots ,t_{L}$ that define the cycles, the number of
iterations $R_{\ell }$ executed in each $M$ phase, and the variance matrices 
$\lambda _{\ell }=\left\{ \Sigma _{\ell r}\right\} $ from each $M$ phase.

\item \label{two_pass_2}Execute algorithm \ref{alg:specific_adaptive} using $%
t_{\ell }$, $R_{\ell }$ and $\lambda _{\ell }$ $\left( \ell =1,\ldots
,L\right) $.
\end{enumerate}

Notice that in step 2 the cycle break points $t_{0},\ldots ,t_{L}$ and the
variance matrices $\Sigma _{\ell r}$ are predetermined with respect to the
particles generated in that step. Because they are fixed with respect to the
process of random particle generation, step \ref{two_pass_2} is a specific
version of Algorithm \ref{alg:general_nonadaptive}. The only change is in
the notation: $\lambda _{\ell }$ in Algorithm \ref{alg:general_nonadaptive}
is the sequence of matrices $\left\{ \Sigma _{\ell r}\right\} $ indexed by $%
r $ in step \ref{two_pass_2} of Algorithm \ref{alg:two_pass}. The results in
Chopin (2004), and other results for SMC algorithms with fixed design
parameters, now apply directly.

The software used for the work in this paper makes it convenient to execute
the two-pass algorithm. In a variety of models and applications results
using Algorithms \ref{alg:specific_adaptive} and \ref{alg:two_pass} have
always been similar, as illustrated in Section \ref{subsec:SMC_reliability}.
Thus it is not necessary to use the two-pass algorithm exclusively, and we
do not recommend doing so in the course of a research project. It is prudent
when SPS is first applied to a new model, because there is no central limit
theorem for the one-pass algorithm (Algorithm \ref{alg:specific_adaptive}),
and one should check early for the possibility that this algorithm might be
inadequate. Given that Algorithm \ref{alg:two_pass} is available in generic
SPS software, and the modest computational cost involved, it is also
probably a wise step in the final stage of research before public
presentation of findings.

\section{Models, data and software\label{sec:Models}}

The balance of this paper studies the performance of the SPS algorithm in a
variety of situations typical of those in applied work. This section
provides full detail of the models used, in Section \ref{subsec:models}, and
describes the data sets used in Section \ref{subsec:data}. The paper
compares the performance of SPS in a variety of software and hardware
environments, and with the state-of-the-art MCMC algorithm described in
Polson et al. (2012). Section \ref{subsec:hardsoftware} provides the details
of the hardware and software used subsequently in Sections \ref%
{sec:SMC_performance} and \ref{sec:comparison} to document the performance
of the PSW and SPS algorithms.

\subsection{\label{subsec:models}Models}

We use the specification (\ref{plogit}) of the multinomial logit model
throughout. The binomial logit model is the special case $C=2$. \ Going
forward, denote the covariates $X=\left[ x_{1},\ldots ,x_{T}\right] ^{\prime
}$. The log odds-ratio%
\begin{equation}
\log \left[ \frac{P\left( Y_{t}=i\mid x_{t},\theta \right) }{P\left(
Y_{t}=j\mid x_{t},\theta \right) }\right] =\left( \theta _{i}-\theta
_{j}\right) ^{\prime }x_{t}\text{\label{log_odds_ratio}}
\end{equation}%
is linear in the parameter vector $\theta $.

While normalization of the parameters is desirable, it is useful to begin
with a Gaussian prior distribution with independent components%
\begin{equation}
\theta _{c}\overset{iid}{\thicksim }N\left( \mu _{c},\Sigma _{c}\right) 
\text{ }\left( c=1,\ldots ,C\right) \text{.\label{prior_Gauss}}
\end{equation}%
This prior distribution implies that the vectors $\theta _{j}-\theta _{c}$ $%
\left( j=1,\ldots ,C;j\not=c\right) $ are jointly normally distributed, with%
\begin{equation}
\mathrm{E}\left( \theta _{j}-\theta _{c}\right) =\mu _{j}-\mu _{c}\text{, }%
\mathrm{var}\left( \theta _{j}-\theta _{c}\right) =\Sigma _{j}+\Sigma _{c}%
\text{, }\mathrm{cov}\left( \theta _{j}-\theta _{c},\theta _{i}-\theta
_{c}\right) =\Sigma _{c}\text{.\label{prior_norm}}
\end{equation}%
This provides the prior distribution of the parameter vector when (\ref%
{plogit}) is normalized by setting $\theta _{c}=0$, that is, when $\theta
_{j}$ is replaced by $\theta _{j}-\theta _{c}$ and $\theta _{c}$ is omitted
from the parameter vector. So long as the constancy of the prior
distribution (\ref{prior_norm}) is respected, all posterior moments of the
form $\mathrm{E}\left[ h\left( Y\right) \mid x\right] $ will be invariant
with respect to normalization. While it is entirely practical to simulate
from the posterior distribution of the unnormalized model, for computation
it is more efficient to use the normalized model because the parameter
vector is shorter, reducing both computing time and storage requirements.

If the prior distribution (\ref{prior_Gauss}) is exchangeable across $%
c=1,\ldots ,C$ then there is no further loss of generality in specifying $%
\mu _{c}=0$ and $\Sigma _{c}=\Sigma $ $\left( c=1,\ldots ,C\right) $. In the
case studies in Section \ref{sec:comparison}, with one minor exception we
restrict $\Sigma $ to the class proposed by Zellner (1986),%
\begin{equation}
\Sigma =g\cdot T\cdot \left( X^{\prime }X\right) ^{-1}\text{.\label%
{g-prior_def}}
\end{equation}%
To interpret $\Sigma $, consider the conceptual experiment in which the
prior distribution of $\theta $ is augmented with $x_{t}$ drawn with
probability $T^{-1}$ from the set $\left\{ x_{1},\ldots ,x_{T}\right\} $ and
then $Y$ is generated by (\ref{plogit}). Then the prior distribution of the
log odds-ratio (\ref{log_odds_ratio}) is also Gaussian, with variance matrix%
\begin{equation*}
\frac{1}{T}\sum_{t=1}^{T}x_{t}^{\prime }\left[ 2gT\left( X^{\prime }X\right)
^{-1}\right] x_{t}=\frac{1}{T}\mathrm{tr}\sum_{t=1}^{T}x_{t}x_{t}^{\prime }%
\left[ 2gT\left( X^{\prime }X\right) ^{-1}\right] =2g\text{.}
\end{equation*}%
The log-odds ratio is centered at 0, with a standard deviation of $\left(
2g\right) ^{1/2}$. Some corresponding 95\% credible sets for the log-odds
ratio are $\left( \log \left( 1/16\right) ,\log \left( 16\right) \right) $
for $g=1$, $\left( \log \left( 1/4\right) ,\log \left( 4\right) \right) $
for $g=1/4$, and $\left( \log \left( 1/2\right) \log \left( 2\right) \right) 
$ for $g=1/16$.

This provides the substantive interpretation of the prior distribution
essential to subjective Bayesian inference.

\subsection{\label{subsec:data}Data}

We used eight different data sets to study and compare the performance of
the PSW and SPS algorithms. \ Table \ref{tab_datasets} summarizes some
properties of these data. The notation in the column headings is taken from
Section \ref{subsec:models}. from which the number of parameters is $k\cdot
\left( C-1\right) $. The values of $g$ in the last column are based on
marginal likelihood approximations, discussed further in Section \ref%
{subsec:comparison_exercise}.

\begin{table}[b] \centering%
\caption{Characteristics of data sets}%
\begin{tabular}{cccccc}
\hline\hline
Data set & Sample size $T$ & Covariates $k$ & Outcomes $C$ & Parameters & 
Modal $g$ \\ \hline
\multicolumn{1}{r}{Diabetes} & 768 & 13 & 2 & 13 & 1/4 \\ 
\multicolumn{1}{r}{Heart} & 270 & 19 & 2 & 19 & 1/4 \\ 
\multicolumn{1}{r}{Australia} & 690 & 35 & 2 & 35 & 1/4 \\ 
\multicolumn{1}{r}{Germany} & 1000 & 42 & 2 & 42 & 1/16 \\ 
\multicolumn{1}{r}{Cars} & 263 & 4 & 3 & 8 & 1/4 \\ 
\multicolumn{1}{r}{Caesarean 1} & 251 & 8 & 3 & 16 & 1/4 \\ 
\multicolumn{1}{r}{Caesarean 2} & 251 & 4 & 3 & 8 & 1 \\ 
\multicolumn{1}{r}{Transportation} & 210 & 9 & 4 & 27 & 1 \\ \hline\hline
\end{tabular}%
\label{tab_datasets}%
\end{table}%

For the binomial logit models, we use the same four data sets as Polson et
al. (2012), Section 3.3. Data and documentation may be found at the
University of California - Irvine Machine Learning Repository\footnote{%
http://archive.ics.uci.edu/ml/datasets.html}, using the links indicated.

\begin{itemize}
\item Data set 1, \textquotedblleft Diabetes.\textquotedblright\ The outcome
variable is indication for diabetes using World Health Organization
criteria, from a sample of individuals of Pima Indian heritage living near
Phoenix, Arizona, USA. Of the covariates, one is a constant and one is a
binary indicator. Link: Pima Indians Diabetes

\item Data set 2, \textquotedblleft Heart.\textquotedblright\ The outcome is
presence of heart disease. Of the covariates, one is a constant and 12 are
binary indicators. $T=270$. Link: Statlog (Heart)

\item Data set 3, \textquotedblleft Australia.\textquotedblright\ The
outcome is approval or denial of an application for a credit card. Of the
covariates, one is a constant and 28 are binary indicators. Link: Statlog
(Australian Credit Approval)

\item Data set 4, \textquotedblleft Germany.\textquotedblright\ The outcome
is approval or denial of credit. Of the covariates, one is a constant and 42
are binary indicators. $T=1000$. Link: Statlog (German Credit Data)
\end{itemize}

For the multinomial logit models, we draw from three data sources. The first
two have been used in evaluating approaches to posterior simulation and the
last is typical of a simple econometric application.

\begin{itemize}
\item Data set 5, \textquotedblleft Cars.\textquotedblright\ The outcome
variable is the kind of car purchased (family, work or sporty). Of the
covariates, one is continuous and the remainder are binary indicators. The
data were used in Scott (2011) in the evaluation of latent variable
approaches to posterior simulation in logit models, and are taken from the
data appendix\footnote{%
http://www-stat.wharton.upenn.edu/\symbol{126}waterman/ fsw/download.html}
of Foster et al. (1998).

\item Data set 6, \textquotedblleft Caesarean 1.\textquotedblright\ The
outcome variable is infection status at birth (none, Type 1, Type 2). The
covariates consist of a constant and three binary indicators. These data
(Farhmeir and Tutz, 2001, Table 1.1) have been a widely used test bed for
the performance posterior simulators given severely unbalanced contingency
tables, for example Fr\"{u}hwirth-Schnatter and Fr\"{u}hwirth (2012) and
references therein. The data are distributed with the $R$ statistical
package.

\item Data set 7, \textquotedblleft Caesarean 2.\textquotedblright\ The data
are the same as in the previous set, except that the model is fully
saturated: there are eight covariates, one for each combination of
indicators. This variant of the model has been widely studied because the
implicit design is severely unbalanced. One cell is empty. For the sole
purpose of constructing the $g$ prior (\ref{g-prior_def}) we supplement the
covariate matrix $X$ with a single row having an indicator in the empty
cell. The likelihood function uses the actual data.

\item Data set 8, \textquotedblleft Transportation.\textquotedblright\ The
data is a choice-based sample of mode of transportation choice (car, bus,
train, air) between Sydney and Melbourne. The covariates are all continuous
except for the intercept. The data (Table F21.2 of the data appendix of
Greene (2003)\footnote{%
http://pages.stern.nyu.edu/\symbol{126}wgreene/Text/tables/tablelist5.htm}
are widely used to illustrate logit choice models in econometrics.
\end{itemize}

\subsection{\label{subsec:hardsoftware}Hardware and software}

The PSW algorithm is described in Polson et al. (2012). The code is the R
package BayesLogit provided by the authors\footnote{%
http://cran.r-project.org/ web/packages/BayesLogit/BayesLogit.pdf}. Except
as noted in Section \ref{subsec:comparison_exercise}, the code executed
flawlessly without intervention. The execution used a 12-core CPU ($2\times $
Intel Xeon 5680) and 24G memory, but the code does not exploit the multiple
cores.

The SPS/CPU algorithm used the algorithm described in Section 2 The code is
written in Matlab Edition 2012b (with the Statistics toolbox) and will be
made available shortly with the next revision of this paper. \ The execution
used a 12-core CPU ($2\times $ Intel Xeon E5645) and 24B memory, exploiting
multiple cores with a ratio of CPU to wall clock time of about 5.\textbf{\ }

The SPS/GPU algorithm used the algorithm described in Section 2. The code is
written Matlab Edition 2012a (with the Statistics and Parallel toolboxes)
and uses the C extension CUDA version 4.2 and will be made available shortly
with the next revision of this paper . The execution used an Intel Core i7
860, 2.8 GHz CPU and one Nvidia GTX 570 GPU\ (480 cores).

\section{Performance of the SPS\ algorithm\label{sec:SMC_performance}}

The SPS algorithm can be used routinely in any model that has a bounded and
directly computed likelihood function, accompanied by a proper prior
distribution. The algorithm performs consistently without intervention from
the user. It provides numerical standard errors that are reliable in the
sense that they indicate correctly the likely outcome of a repeated,
independent execution of the sequential posterior simulator. As a
by-product, it also provides consistent (in $N$) approximations of log
marginal likelihood and associated numerical standard error; Section 4 of
Durham and Geweke (2012) explains the procedure. Section \ref%
{subsec:SMC_reliability}, below, illustrates these properties for the case
of the multinomial logit model. The frequency of transition from one cycle
to a new cycle as well as the number of steps taken in the $M$ phase, and
therefore the execution time, depend on characteristics of the model and the
data. Section \ref{subsec:SMC_adaptation} studies some aspects of this
dependence for the case of the multinomial logit model.

\subsection{\label{subsec:SMC_reliability}Reliability of the SPS algorithm}

Numerical approximations of posterior moments must be accompanied by an
indication of their accuracy. Even if editorial constraints make it
impossible to accompany each moment approximation with an indication of
accuracy, decent scholarship demands that the investigator be aware of the
accuracy of reported moment approximations. Moreover, the accuracy
indications must themselves be interpretable and reliable.

The SPS methodology for the logit model described in Section \ref{sec:SMC}
achieves this standard by means of a central limit theorem for posterior
moment approximations accompanied by a scheme for grouping particles that
leads to a simple simulation-consistent approximation of the variance in the
central limit theorem. The practical manifestation of these results is the
numerical standard error (\ref{NSE_def}). The underlying theory for SPS
requires the two-pass procedure of Algorithm \ref{alg:two_pass}. If the
theory is adequate, then numerical standard errors form the basis for
reliable predictions of the outcome of repeated simulations.

\begin{table}[tb] \centering%
\caption{Reliability of one- and two-pass algorithms}%
\begin{tabular}{llll}
\hline\hline
& $\mathrm{E}\left( \theta _{1}^{\prime }\overline{x}\mid \mathrm{Data}%
\right) $ & $\mathrm{E}\left( \theta _{2}^{\prime }\overline{x}\mid \mathrm{%
Data}\right) $ & $\log $ ML \\ \hline
$J=10,N=1000$ &  &  &  \\ 
Run $A$, Pass 1 & 0.6869 [.0017] & -0.3836 [.0017] & -253.732 [.057] \\ 
Run $A$, Pass 2 & 0.6855 [.0012] & -0.3856 [.0024] & -253.504 [.086] \\ 
Run $B$, Pass 1 & 0.6811 [.0014] & -0.3920 [.0017] & -253.522 [.082] \\ 
Run $B$, Pass 2 & 0.6847 [.0018] & -0.3877 [.0025] & -253.514 [.054] \\ 
Run $C$, Pass 1 & 0.6844 [.0016] & -0.3892 [.0018] & -253.478 [.055] \\ 
Run $C$, Pass 2 & 0.6837 [.0019] & -0.3875 [.0021] & -253.637 [.065] \\ 
\hline
$J=40,N=2500$ &  &  &  \\ 
Run $A$, Pass 1 & 0.6850 [.00051] & -0.3873 [.00062] & -253.6117 [.026] \\ 
Run $A$, Pass 2 & 0.6857 [.00049] & -0.3871 [.00076] & -253.5934 [.029] \\ 
Run $B$, Pass 1 & 0.6844 [.00043] & -0.3890 [.00061] & -253.6249 [.019] \\ 
Run $B$, Pass 2 & 0.6849 [.00048] & -0.3885 [.00063] & -253.5708 [.027] \\ 
Run $C$, Pass 1 & 0.6853 [.00050] & -0.3881 [.00060] & -253.5918 [.023] \\ 
Run $C$, Pass 2 & 0.6847 [.00043] & -0.3872 [.00052] & -253.5922 [.024] \\ 
\hline\hline
\end{tabular}%
\label{tab_reliable}%
\end{table}%

Table \ref{tab_reliable} provides some evidence on these points using the
multinomial logit model and the cars data set described in Section \ref%
{subsec:data}. For both small and large SPS executions (upper and lower
panels, respectively) Table \ref{tab_reliable} indicates moment
approximations for three independent executions ($A$, $B$ and $C$) of the
two-pass algorithm, and for both the first and second pass of the algorithm.
The posterior moments used in the illustrations here, and subsequently, are
the the log-odds (with respect to the outcome $Y=C$, $\theta _{c}^{\prime }%
\overline{x}$ $\left( c=1,\ldots ,C\right) $, where $\overline{x}$ is the
sample mean of the covariates. Numerical standard errors are indicated in
brackets. As discussed in Section \ref{subsec:Adaptive}, these will vary
quite a bit more from one run to another when $J=10$ than they will when $%
J=40$, and this is evident in Table \ref{tab_reliable}.

Turning first to the comparison of results at the end of Pass 1 (no formal
justification for numerical standard errors) and at the end of Pass 2 (the
established results for the nonadaptive algorithm discussed in Section \ref%
{subsec:Nonadaptive} apply) there are no unusually large differences within
any run, given the numerical standard errors. That is, there is no evidence
to suggest that if an investigator used Pass 1 results to anticipate what
Pass 2 results would be, the investigator would be misled.

This still leaves the question of whether the numerical standard errors from
a single run are a reliable indication of what the distribution of Monte
Carlo approximations would be across independent runs. Comparing results for
runs $A$, $B$ and $C$, Table \ref{tab_reliable} provides no indication of
difficulty for the large SPS executions. For the small SPS executions, there
is some suggestion that variation across runs at the end of the first pass
is larger than numerical standard error suggests. Note in particular the
approximation of log marginal likelihood for runs $A$ and $C$, and note also 
$E\left( \theta _{1}^{\prime }\overline{x}\right) $ for runs $A$ and $B$.

These suggestions could be investigated more critically with scores or
hundreds of runs of the SPS algorithm, but we conjecture the returns would
be low and in any event there is no basis for extrapolating results across
models and data sets. Most important, one cannot resort to this tactic
routinely in applied work. The results here support the earlier
recommendation (at the end of Section \ref{sec:Models_etc}) that the
investigator proceed mainly using the one-pass algorithm, reserving the
two-pass algorithm for checks at the start and the end of a research project.

\subsection{\label{subsec:SMC_adaptation}Adaptation in the SPS algorithm}

The SPS algorithm approximates posterior distributions by mimicking the
formal Bayesian updating process, observation by observation, using (at
least) thousands of particles simultaneously. It does so in a reliable and
robust manner, with much less intervention, problem-specific tailoring, or
baby-sitting than is the case with other posterior simulation methods. For
example it does not require the investigator to tailor a source distribution
for importance sampling (which SPS uses in the $C$ phase) nor does it
require that the investigator monitor a Markov chain (which SPS uses in the $%
M$ phase) for stationarity or serial correlation.

\begin{table}[tb] \centering%
\caption{Some features of the SMC algorithm for the cars data}%
\begin{tabular}{llllll}
\hline\hline
& $g=1/64$ & $g=1/16$ & $g=1/4$ & $g=1$ & $g=4$ \\ \hline
$J=10,N=1000$ &  &  &  &  &  \\ 
Cycles & 11 & 18 & 24 & 32 & 37 \\ 
M iterations & 80 & 186 & 189 & 257 & 330 \\ 
Relative time & 1.00 & 2.08 & 1.47 & 2.11 & 2.19 \\ 
RNE, $\mathrm{E}\left( \theta _{1}^{\prime }\overline{x}\mid \mathrm{Data}%
\right) $ & 1.08 & 1.09 & 1.02 & 1.55 & 1.06 \\ 
RNE, $\mathrm{E}\left( \theta _{2}^{\prime }\overline{x}\mid \mathrm{Data}%
\right) $ & 0.93 & 0.86 & 0.83 & 2.15 & 2.13 \\ 
NSE(log ML) & 0.088 & 0.052 & 0.069 & 0.075 & 0.089 \\ \hline
$J=40,N=2500$ &  &  &  &  &  \\ 
Cycles & 11 & 18 & 24 & 33 & 37 \\ 
M iterations & 94 & 131 & 178 & 268 & 319 \\ 
Relative time & 1.00 & 1.09 & 1.32 & 1.76 & 1.92 \\ 
RNE, $\mathrm{E}\left( \theta _{1}^{\prime }\overline{x}\mid \mathrm{Data}%
\right) $ & 1.00 & 0.94 & 1.12 & 1.17 & 1.03 \\ 
RNE, $\mathrm{E}\left( \theta _{2}^{\prime }\overline{x}\mid \mathrm{Data}%
\right) $ & 1.12 & 0.99 & 1.19 & 1.23 & 1.01 \\ 
NSE(log ML) & 0.025 & 0.027 & 0.026 & 0.035 & 0.037 \\ \hline\hline
\end{tabular}%
\label{tab_technical}%
\end{table}%

\begin{figure}
\centering
\includegraphics[width=6.0502in, height=3.8532in]{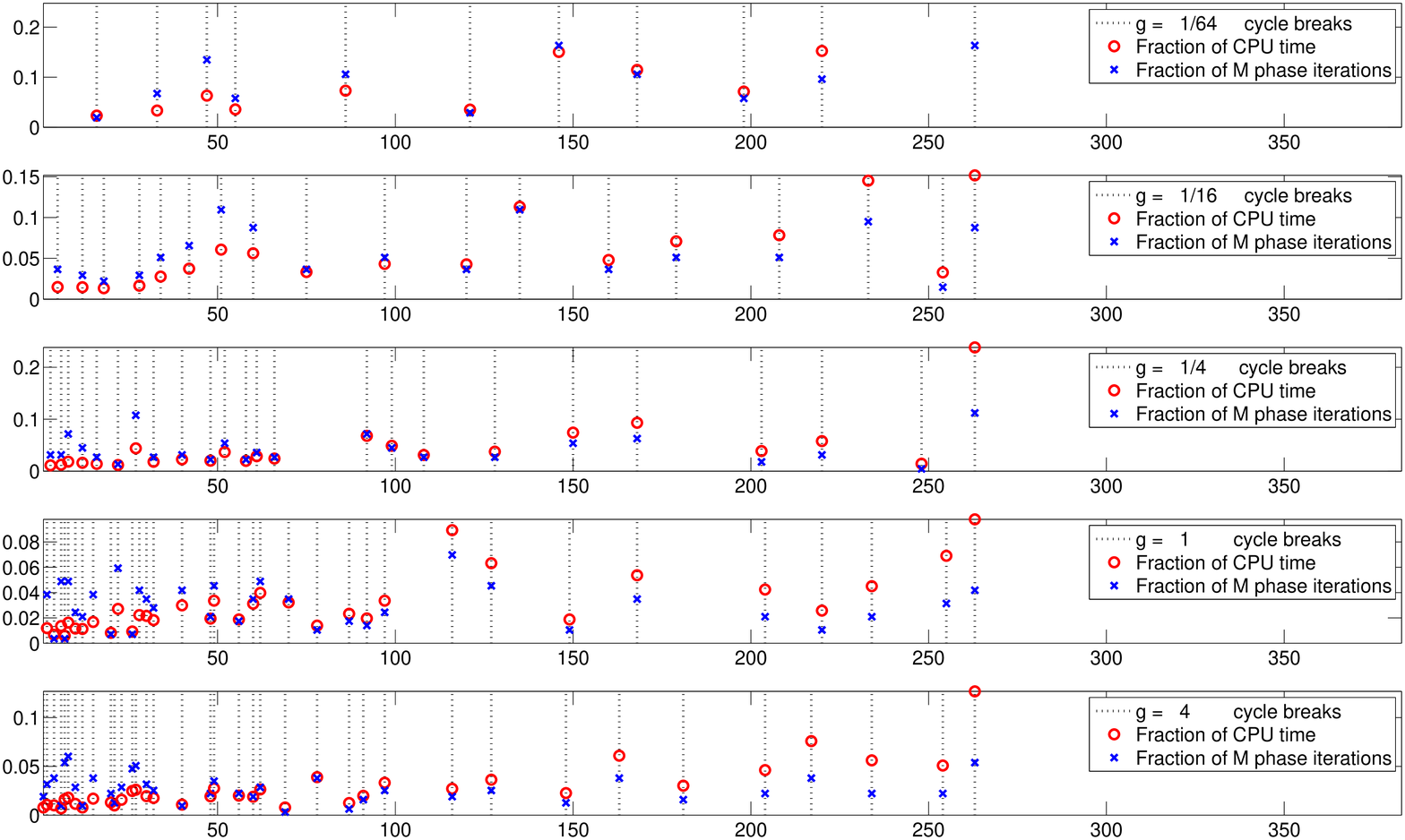}
\caption{Some properties of the SMC algorithm in the cars example $\left( J=10,N=1000\right) $}
\label{fig_cycles_10K}
\end{figure}

\begin{figure}
\centering
\includegraphics[width=6.0502in, height=3.8532in]{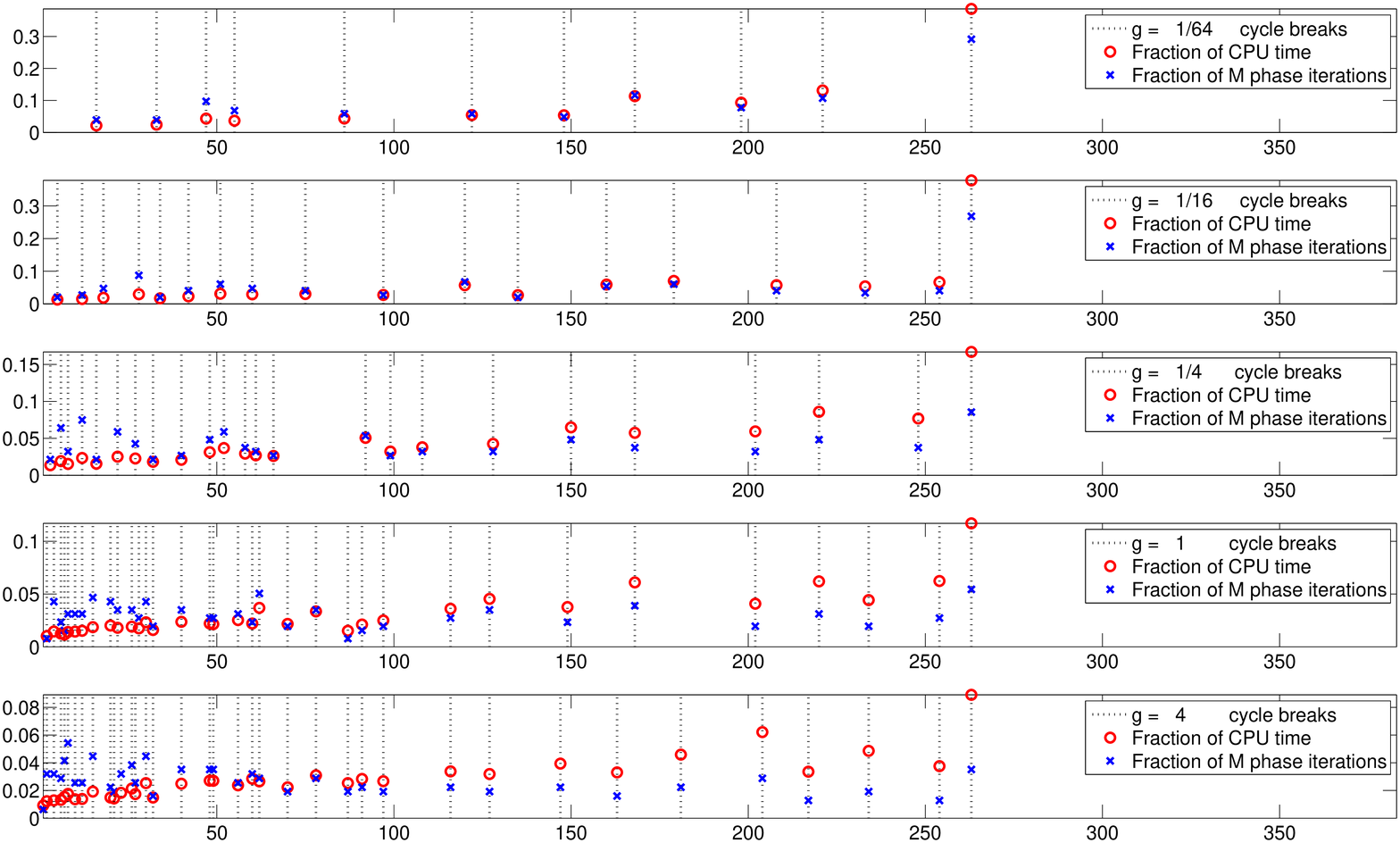}
\caption{Some properties of the SMC algorithm in the cars example $\left( J=40,N=2500\right) $}
\label{fig_cycles_100K}
\end{figure}

While SPS requires very little intervention by the user, a little insight
into its mechanics helps to understand the computational demands of the
algorithm. Table \ref{tab_technical} and Figures \ref{fig_cycles_10K} and %
\ref{fig_cycles_100K} break out some details of these mechanics, continuing
to use the cars data set from Section \ref{subsec:SMC_reliability}. Figure %
\ref{fig_cycles_10K} and the upper panel of Table \ref{tab_technical}
pertain to the small SPS execution, and Figure \ref{fig_cycles_100K} and the
lower panel of Table \ref{tab_technical} pertain to the large SPS execution.
They compare performance under all five prior distributions to illustrate
some central features of the algorithm.

Essentially by design, the algorithm achieves similar accuracy of
approximation for all five prior distributions. For moments, this is driven
by the iterations in the final $M$ phase that terminate only when the RNE
for all monitoring functions first exceeds 0.9. The monitoring functions are
not the same as the log-odds ratio functions of interest, and log marginal
likelihood is not a posterior moment, but the principle that computation
goes on until a prescribed criterion of numerical accuracy is achieved is
common to all models and applications. Relative numerical efficiencies show
less variation across the five cases for the large SPS executions for the
usual reason: one learns more about reliability from $J=40$ particle groups
than from $J=10$ particle groups.

The most striking feature of Table 2 is that more diffuse prior
distributions (e.g., $g=1$, $g=4$) lead to more cycles and total iterations
in the $M$ phase than do tighter prior distributions (e.g. $g=1/64$, $g=1/16$%
). Indeed, the ratio of $M$ phase iterations to cycles remains roughly
constant. The key to understanding this behavior is the insight that the
algorithm terminates the addition of information in the $C$ phase, thus
defining a cycle, when the accumulation of new information has introduced
enough variation in particle weights that effective sample size drops below
the threshold of half the number of particles.

Figures \ref{fig_cycles_10K} and \ref{fig_cycles_100K} show the cycle
breakpoints under each of the five prior distributions. Breakpoints tend to
be more frequent near the start of the sample and algorithm, when the
relative contribution of each observation to the posterior distribution is
greatest. This is true in all five cases. As the prior distribution becomes
more diffuse the posterior becomes more sensitive to each observation. This
sensitivity is concentrated at the start of the sample and algorithm.

Later in the sample changes in the weight function are driven more by
particular observations that contribute more information. These are the
observations that are less likely conditional on previous observations,
contributing to greater changes in the posterior distribution and therefore
increasing the variation in particle weights and triggering new cycles. This
is evident in breakpoints that are the same under all five prior
distributions.

Consequently, the additional cycles and breakpoints arising from more
diffuse priors tend to be concentrated in the earlier part of the algorithm.
Sample sizes are smaller here than later, and the repeated evaluations of
the likelihood function in the $M$ phase demand fewer computations. This is
the main reason that relative computation time increase by a factor of less
than 2, in moving from the tightest to the most diffuse prior in Table \ref%
{tab_technical} , whereas the number of cycles and $M$ phase iterations more
than triples.

The limiting cases are simple and instructive. A dogmatic prior implies a
uniform weight function and one cycle. For most likelihood functions, in the
limit a sequence of increasingly diffuse priors guarantees $t_{1}=1$, the
existence exactly one unique particle in each group at the conclusion of the
first $S$ phase, and therefore $\mathrm{rank}\left( V_{11}\right) =\min
\left( J,\dim \left( \theta \right) \right) $ in the $M$ phase, and the
algorithm will fail at step \ref{failstep}.

While the theory requires only that the prior distribution be proper, the
SPS algorithm functions best for prior distributions that are seriously
subjective -- for example, the prior distribution developed in Section \ref%
{subsec:models}. This requirement arises from the representation of the
Bayesian updating procedure by means of a finite number of points. Our
experience in this and other models is that a proper prior distribution with
a reasoned substantive interpretation presents no problems for the SPS
algorithm. The next section illustrates this point.

\section{\label{sec:comparison}Comparison of algorithms}

We turn now to a systematic comparison of the efficiency and reliability of
the PSW and SPS algorithms in the logit model.

\subsection{The exercise\label{subsec:comparison_exercise}}

To this end, we simulated the posterior distribution for the Cartesian
product of the eight data sets described in Section \ref{subsec:data} and
Table \ref{tab_datasets}, the five prior distributions utilized in Section %
\ref{subsec:SMC_adaptation} and Table \ref{tab_technical}, and five
approaches to simulation. The first approach to posterior simulation is the
PSW algorithm implemented as described in Section \ref{subsec:hardsoftware}.
The second approach uses the small SPS simulation ($J=10$, $N=1000$) with
the CPU implementation described in Section \ref{subsec:hardsoftware}, and
the third approach uses the GPU implementation. The fourth and fifth
approaches are the same as the second and third except that they use the
large SPS simulation ($J=40$, $N=2500$).

To complete this exercise we had to modify the multinomial logit model (last
four data sets) for the PSW algorithm. The code that accompanies the
algorithm requires that the vectors $\theta _{c}$ be independent in the
prior distribution, and consequently (\ref{g-prior_def}) was modified to
specify $\mathrm{cov}\left( \theta _{j}-\theta _{c},\theta _{i}-\theta
_{c}\right) =0$. As a consequence, posterior moments approximated by the PSW
algorithm depend on the normalization employed and are never the same as
those approximated by the SPS algorithms. We utilized the same normalization
as in the SPS algorithms, except for the cars data set, for which the code
would not execute with this choice and we normalized instead on the second
choice.

\begin{table}[tb] \centering%
\caption{Log marginal likelihoods, all data sets and models}%
\begin{tabular}{llllll}
\hline\hline
& $g=1/64$ & $g=1/16$ & $g=1/4$ & $g=1$ & $g=4$ \\ \hline
\multicolumn{1}{r}{Diabetes} & -405.87 [0.04] & -386.16 [0.03] & -383.31
[0.03] & -387.01 [0.04] & -392.61 [0.04] \\ 
\multicolumn{1}{r}{Heart} & -141.00 [0.04] & -123.36 [0.03] & -118.58 [0.04]
& -124.38 [0.06] & -135.25 [0.11] \\ 
\multicolumn{1}{r}{Australia} & -301.87 [0.04] & -269.90 [0.05] & -267.41
[0.06] & -280.47 [0.07] & -300.20 [0.12] \\ 
\multicolumn{1}{r}{Germany} & -539.46 [0.06] & -535.91 [0.08] & -556.71
[0.00] & -586.66 [0.00] & -621.53 [0.00] \\ 
\multicolumn{1}{r}{Cars} & -263.75 [0.02] & -254.42 [0.03] & -253.62 [0.03]
& -257.18 [0.03] & -262.20 [0.04] \\ 
\multicolumn{1}{r}{Caesarean 1} & -214.50 [0.03] & -187.19 [0.03] & -176.96
[0.02] & -177.29 [0.03] & -181.66 [0.03] \\ 
\multicolumn{1}{r}{Caesarean 2} & -219.20 [0.03] & -192.10 [0.03] & -180.30
[0.02] & -178.91 [0.02] & -181.42 [0.02] \\ 
\multicolumn{1}{r}{Transportation} & -234.58 [0.03] & -197.07 [0.04] & 
-176.14 [0.05] & -173.97 [0.05] & -184.24 [0.07] \\ \hline\hline
\end{tabular}%
\label{tab_logMLs}%
\end{table}%

Since it is impractical to present results from all $8\times 5\times 5=200$
posterior simulations, we restrict attention to a single prior distribution
for each data set: the one producing the highest marginal likelihood. Table
4 provides the log marginal likelihoods under all five prior distributions
for all eight data sets, as computed using the large SPS/GPU algorithm.

Note that the accuracy of these approximations is very high, compared with
existing standards for posterior simulation. The accuracy of log-marginal
likelihood approximation tends to decline with increasing sample size, as
detailed in Durham and Geweke (2012, Section 4) and this is evident in Table %
\ref{tab_logMLs}. Going forward, all results pertain to the g-prior
described in Section 3.1 with $g=1/16$ for Germany, $g=1$ for Caesarean 1
and transportation, and $g=1/4$ for the other five data sets.

\begin{table}[tb] \centering%
\caption{Posterior moments and numerical accuracy}%
\begin{tabular}{cccccc}
\hline\hline
&  & \multicolumn{2}{c}{$(J=10,N=1000)$} & \multicolumn{2}{c}{$J=40,N=2500)$}
\\ 
& PSW & SPS/CPU & SPS/GPU & SPS/CPU & SPS/GPU \\ \hline
\multicolumn{1}{r}{Diabetes} & -0.853 (.096) & -0.855 (.095) & -0.852 (.096)
& -0.853 (.095) & -0.853 (.095) \\ 
\multicolumn{1}{r}{} & [.0008, 0.68] & [.0009, 1.12] & [.0009, 1.21] & 
[.0003, 0.99] & [.0003, 1.03] \\ 
\multicolumn{1}{r}{Heart} & -0.250 (.192) & -0.246 (.187) & -0.251 (.191) & 
-0.249 (.189) & -0.250 (.189) \\ 
\multicolumn{1}{r}{} & [.0021, 0.43] & [.0019, 0.96] & [.0019, 0.98] & 
[.0006, 0.86] & [.0006, 0.92] \\ 
\multicolumn{1}{r}{Australia} & -0.438, .157) & -0.438 (.157) & -0.439 (.156)
& -0.440 (.156) & -0.438 (.157) \\ 
\multicolumn{1}{r}{} & [.0023, 0.23] & [.0016, 0.96] & [.0016, 0.98] & 
[.0005, 0.97] & [.0006, 0.60] \\ 
\multicolumn{1}{r}{Germany} & -1.182 (.089) & -1.180 (.089) & -1.81 (.089) & 
-1.182 (.089) & -1.81 (.088) \\ 
\multicolumn{1}{r}{} & [.0010, 0.36] & [.0009, 1.00] & [.0004, 0.94] & 
[.0003, 0.91] & [.0004, 0.94] \\ 
\multicolumn{1}{r}{Cars} & -1.065 (.171) & 0.684 (.156) & 0.685 (.158) & 
0.685 (.156) & 0.685 (.156) \\ 
\multicolumn{1}{r}{} & [.0002, 0.46] & [.0015, 1.03] & [.0017, 0.98] & 
[.0005, 1.12] & [.0005, 0.97] \\ 
\multicolumn{1}{r}{} & -0.665 (.156) & -0.386 (.195) & -0.388 (.195) & 
-0.387 (.194) & -0.388 (.193) \\ 
\multicolumn{1}{r}{} & [.0009, 0.60] & [.0021, 0.83] & [.0017, 1.29] & 
[.0006, 1.19] & [.0007, 0.72] \\ 
\multicolumn{1}{r}{Caesarean 1} & -1.975 (.241) & -2.049 (.245) & -2.052
(.248) & -2.052 (.246) & -2.052 (.246) \\ 
\multicolumn{1}{r}{} & [.0004, 0.26] & [.0024, 1.09] & [.0037, 0.45] & 
[.0008, 0.91] & [.0008, 0.96] \\ 
\multicolumn{1}{r}{} & -1.607 (.211) & -1.698 (.215) & -1.694 (.217) & 
-1.697 (.219) & -1.698 (.219) \\ 
\multicolumn{1}{r}{} & [.0003, 0.27] & [.0018, 1.36] & [.0024, 0.80] & 
[.0007, 1.07] & [.0006, 1.30] \\ 
\multicolumn{1}{r}{Caesarean 2} & -2.033 (.261) & -2.057 (.264) & -2.056
(.264) & -2.056 (.262) & -2.0534 (.262) \\ 
\multicolumn{1}{r}{} & [.0004, 0.22] & [.0027, 0.94] & [.0025, 1.32] & 
[.0008, 0.99] & [.0009, 0.89] \\ 
\multicolumn{1}{r}{} & -1.586 (.205) & -1.597 (.210) & -1.587 (.206) & 
-1.593 (.206) & -1.593 (.207) \\ 
\multicolumn{1}{r}{} & [.0003, 1.30] & [.0021, 0.98] & [.0021, 0.99] & 
[.0006, 1.02] & [.0006, 1.03] \\ 
\multicolumn{1}{r}{Transportation} & 0.091 (.316) & 0.130 (.321) & 0.119
(.322) & 0.123 (.322) & 0.123 (.323) \\ 
\multicolumn{1}{r}{} & [.0006, 0.13] & [.0031, 1.09] & [.0030, 1.14] & 
[.0010, 1.01] & [.0010, 0.97] \\ 
\multicolumn{1}{r}{} & -0.588 (.400) & -0.416 (.380) & -0.416 (.388) & 
-0.421 (.386) & -0.419 (.388) \\ 
\multicolumn{1}{r}{} & [.0009, 0.09] & [.0020, 3.52] & [.0043, 0.82] & 
[.0012, 1.04] & [.0011, 1.29] \\ 
\multicolumn{1}{r}{} & -1.915 (.524) & -1.646 (.487) & -1.642 (.490) & 
-1.645 (.491) & -1.647 (.492) \\ 
\multicolumn{1}{r}{} & [.0013, 0.07] & [.0036, 1.84] & [.0044, 1.24] & 
[.0016, 0.95] & [.0019, 0.65] \\ \hline\hline
\end{tabular}%
\label{tab_moments}%
\end{table}%

\subsection{Reliability}

We assess the reliability of the algorithms by comparing posterior moment
and marginal likelihood approximations for the same model. Table \ref%
{tab_moments} provides the posterior moment approximations. The moment used
is, again, the posterior expectation of the log-odds ratio(s) evaluated at
the sample mean $\overline{x}$, for each choice relative to the last choice.
(This corresponds to the normalization used in execution.) Thus there is one
moment for each of the four binomial logit data sets, two moments for the
first three multinomial logit data sets, and three moments for the last
multinomial logit model data set.

The result for each moment and algorithm is presented in a block of four
numbers. The first line has the simulation approximation of the posterior
expectation followed by the simulation approximation of the posterior
standard deviation. The second line contains [in brackets] the numerical
standard error and relative numerical efficiency of the approximation. For
the multinomial logit model there are multiple blocks, one for each
posterior moment.

For the SPS algorithms the numerical standard error and relative numerical
efficiency are the natural by-product of the results across the $J$ groups
of particles as described in Section \ref{subsec:Conditions}. For the PSW
algorithm these are computed based on the 100 independent executions of the
algorithm. The PSW approximations of the posterior expectation and standard
deviation are based on a single execution.

Posterior moment approximations are consistent across algorithms. For the
PSW algorithms there are $6\times 13=78$ pairwise comparisons of posterior
expectations that can be made, and of these two are in the upper 1\% or
lower 1\% tail of the distribution. The PSW moment approximations are
consistent with the SPS moment approximations for the binomial logit data
sets. As explained earlier in this section, the moments approximated by the
PSW algorithm are not exactly the same as those approximated by the SPS
algorithms in the last four data sets.

Table \ref{tab_logMLs2} compares approximations of log marginal likelihoods
across the four variants of the SPS algorithm, and there are no anomalies.
(The PSW algorithm does not yield approximations of the marginal
likelihood.) There is no evidence of unreliability of any of the algorithms
in Tables \ref{tab_moments} and \ref{tab_logMLs2}.

\begin{table}[tb] \centering%
\caption{Log marginal likelihoods and numerical accuracy}%
\begin{tabular}{lllll}
\hline\hline
& \multicolumn{2}{c}{$(J=10,N=1000)$} & \multicolumn{2}{c}{$J=40,N=2500)$}
\\ 
& SPS/CPU & SPS/GPU & SPS/CPU & SPS/GPU \\ \hline
\multicolumn{1}{r}{Diabetes} & -383.15 [0.05] & -383.14 [0.17] & -383.31
[0.03] & -383.25 [0.03] \\ 
\multicolumn{1}{r}{Heart} & -118.29 [0.15] & -118.73 [0.14] & -118.58 [0.04]
& -118.61 [0.04] \\ 
\multicolumn{1}{r}{Australia} & -267.25 [0.32] & -267.35 [0.19] & -267.41
[0.06] & -267.35 [0.05] \\ 
\multicolumn{1}{r}{Germany} & -536.05 [0.21] & -536.10 [0.18] & -535.91
[0.08] & -535.89 [0.07] \\ 
\multicolumn{1}{r}{Cars} & -253.57 [0.07] & -253.46 [0.10] & -253.62 [0.03]
& -253.61 [0.03] \\ 
\multicolumn{1}{r}{Caesarean 1} & -177.06 [0.08] & -176.72 [0.13] & -176.96
[0.02] & -176.91 [0.03] \\ 
\multicolumn{1}{r}{Caesarean 2} & -178.89 [0.08] & -178.95 [0.03] & -178.91
[0.02] & -178.83 [0.03] \\ 
\multicolumn{1}{r}{Transportation} & -174.09 [0.12] & -173.92 [0.19] & 
-173.97 [0.05] & -173.99 [0.05] \\ \hline\hline
\end{tabular}%
\label{tab_logMLs2}%
\end{table}%

\subsection{Computational efficiency}

Our comparisons are based on a single run of each of the five algorithms
(PSW and four variants of SPS) for each of the eight data sets, using for
each data set one particular prior distribution chosen as indicated in
Section \ref{subsec:comparison_exercise}. In the case of the PSW algorithm,
we used 20,000 iterations for posterior moment approximation for the first
four data sets, and 21,000 for the latter four data sets. The entries show
wall-clock time for execution on the otherwise idle machine described in
Section \ref{subsec:hardsoftware}. \ Execution time for the PSW algorithm
1,000 burn-in iterations in all cases except Australia and Germany, which
have 5,000 burn-in iterations.\ Times can very considerably, depending on
the particular hardware used: for example, the SPS/CPU algorithms were
executed using a 12-core machine that utilized about 5 cores,
simultaneously, on average; and the SPS/GPU algorithms used only a single
GPU with 480 cores. The results here must be qualified by these
considerations. We suspect that in practice the SPS/CPU algorithm might be
slower for many users who have fewer CPU cores; and the SPS/GPU algorithm
might be considerably faster with more GPUs.

\begin{table}[tb] \centering%
\caption{Clock execution time}%
\begin{tabular}{llllll}
\hline\hline
&  & \multicolumn{2}{c}{$(J=10,N=1000)$} & \multicolumn{2}{c}{$J=40,N=2500)$}
\\ 
& PSW & SPS/CPU & SPS/GPU & SPS/CPU & SPS/GPU \\ \hline
\multicolumn{1}{r}{Diabetes} & \multicolumn{1}{r}{14.90} & 
\multicolumn{1}{r}{106.7} & \multicolumn{1}{r}{6.00} & \multicolumn{1}{r}{
739.9} & \multicolumn{1}{r}{26.9} \\ 
\multicolumn{1}{r}{Heart} & \multicolumn{1}{r}{9.53} & \multicolumn{1}{r}{
140.8} & \multicolumn{1}{r}{13.7} & \multicolumn{1}{r}{923.4} & 
\multicolumn{1}{r}{73.6} \\ 
\multicolumn{1}{r}{Australia} & \multicolumn{1}{r}{41.60} & 
\multicolumn{1}{r}{1793.5} & \multicolumn{1}{r}{69.2} & \multicolumn{1}{r}{
12449.9} & \multicolumn{1}{r}{448.7} \\ 
\multicolumn{1}{r}{Germany} & \multicolumn{1}{r}{125.59} & 
\multicolumn{1}{r}{5910.4} & \multicolumn{1}{r}{225.9} & \multicolumn{1}{r}{
45263.2} & \multicolumn{1}{r}{1689,4} \\ 
\multicolumn{1}{r}{Cars} & \multicolumn{1}{r}{7.62} & \multicolumn{1}{r}{33.5
} & \multicolumn{1}{r}{3.5} & \multicolumn{1}{r}{231.2} & \multicolumn{1}{r}{
18.9} \\ 
\multicolumn{1}{r}{Caesarean 1} & \multicolumn{1}{r}{6.84} & 
\multicolumn{1}{r}{97.3} & \multicolumn{1}{r}{10.9} & \multicolumn{1}{r}{
723.3} & \multicolumn{1}{r}{55.3} \\ 
\multicolumn{1}{r}{Caesarean 2} & \multicolumn{1}{r}{6.65} & 
\multicolumn{1}{r}{15.2} & \multicolumn{1}{r}{2.5} & \multicolumn{1}{r}{133.3
} & \multicolumn{1}{r}{11.6} \\ 
\multicolumn{1}{r}{Transportation} & \multicolumn{1}{r}{15.10} & 
\multicolumn{1}{r}{569.7} & \multicolumn{1}{r}{39.7} & \multicolumn{1}{r}{
3064.2} & \multicolumn{1}{r}{293.7} \\ \hline\hline
\end{tabular}%
\label{tab_runtimes}%
\end{table}%

Execution time also depends on memory management, clearly evident in Table %
\ref{tab_runtimes}. The ratio of execution time for the SPS/CPU algorithm in
the large simulations ($J=40,N=2500$) \ to that in the small simulations ($%
J=10,N=1000$) ranges from from 8.5 (Transportation) to 16.2 (Australia). \
There is no obvious pattern or source for this variation, which we are
investigating further. \ The same ratio for the SPS/GPU algorithm ranges
from 4.48 (Diabetes) to about 7.45 (Germany and Transportation). This
reflects the fact that GPU\ computing is more efficient to the extent that
the application is intensive in arithmetic logic as opposed to flow control.
Very small problems are relatively inefficient; as the number and size of
particles increases, the efficiency of the SPS/GPU algorithm increases,
approaching an asymptotic ratio of number and size of particles to computing
time from below.

Relevant comparisons of computing time $t$ require that we correct for the
number $\widetilde{M}$ of iterations or particles and the relative numerical
efficiency $R\widetilde{N}E$ of the algorithm. This produces an
efficiency-adjusted computing time $\widetilde{t}=t/\left( \widetilde{M}%
\cdot R\widetilde{N}E\right) $. For $R\widetilde{N}E\ $we use the average of
the relevant RNE's reported in Table \ref{tab_moments}: in the case of SPS,
the averages are taken across all four variants since population RNE does
not depend on the number of particles, hardware or software. This ignores
variation in RNE from moment to moment and one run to the next. In the case
of PSW, it also ignores dependence of RNE and number of burn-in iterations
on the number of iterations used for moment approximations that arises from
both practical and theoretical considerations. Therefore efficiency
comparisons should be taken as indicative rather than definitive: they will
vary from application to application in any event, and one will not
undertake these comparisons for every (if indeed any) substantive study.

Table \ref{tab_efficiencies} provides the ratio of $\widetilde{t}$ for each
of the SPS algorithms to $\widetilde{t}$ for the PSW\ algorithm, for each of
the eight data sets. The SPS/CPU algorithm compares more favorably with the
PSW algorithm for the small simulation exercises than for the large
simulation exercises. \ The SPS/CPU algorithm is clearly slower than the PSW
algorithm, and its disadvantage becomes more pronounced the greater the
number of parameters and observations. \ With a single exception (Germany)
the SPS/GPU algorithm is faster than the PSW algorithm for the large
simulation exercises, and for the single exception it is about 2\% slower.

\begin{table}[tb] \centering%
\caption{Computational efficiency relative to PSW}%
\begin{tabular}{lllll}
\hline\hline
& \multicolumn{2}{c}{$(J=10,N=1000)$} & \multicolumn{2}{c}{$J=40,N=2500)$}
\\ 
& SPS/CPU & SPS/GPU & SPS/CPU & SPS/GPU \\ \hline
\multicolumn{1}{r}{Diabetes} & \multicolumn{1}{r}{9.40} & \multicolumn{1}{r}{
0.53} & \multicolumn{1}{r}{6.52} & \multicolumn{1}{r}{0.24} \\ 
\multicolumn{1}{r}{Heart} & \multicolumn{1}{r}{14.35} & \multicolumn{1}{r}{
1.40} & \multicolumn{1}{r}{9.41} & \multicolumn{1}{r}{0.75} \\ 
\multicolumn{1}{r}{Australia} & \multicolumn{1}{r}{28.25} & 
\multicolumn{1}{r}{1.09} & \multicolumn{1}{r}{19.61} & \multicolumn{1}{r}{
0.71} \\ 
\multicolumn{1}{r}{Germany} & \multicolumn{1}{r}{45.06} & \multicolumn{1}{r}{
1.72} & \multicolumn{1}{r}{34.51} & \multicolumn{1}{r}{1.29} \\ 
\multicolumn{1}{r}{Cars} & \multicolumn{1}{r}{5.04} & \multicolumn{1}{r}{0.53
} & \multicolumn{1}{r}{3.47} & \multicolumn{1}{r}{0.28} \\ 
\multicolumn{1}{r}{Caesarean 1} & \multicolumn{1}{r}{8.36} & 
\multicolumn{1}{r}{0.94} & \multicolumn{1}{r}{6.21} & \multicolumn{1}{r}{0.47
} \\ 
\multicolumn{1}{r}{Caesarean 2} & \multicolumn{1}{r}{3.73} & 
\multicolumn{1}{r}{0.62} & \multicolumn{1}{r}{3.28} & \multicolumn{1}{r}{0.29
} \\ 
\multicolumn{1}{r}{Transportation} & \multicolumn{1}{r}{6.19} & 
\multicolumn{1}{r}{0.43} & \multicolumn{1}{r}{3.33} & \multicolumn{1}{r}{0.32
} \\ \hline\hline
\end{tabular}%
\label{tab_efficiencies}%
\end{table}%

\section{Conclusion}

The class of sequential posterior simulation algorithms is becoming an
important subset of the computational tools that Bayesian statisticians have
at their disposal in applied work. Graphical processing units, in turn, have
become a convenient and very cost-effective platform for scientific
computing, potentially accelerating computing speeds by orders of magnitude
for suitable algorithms. One of the appealing features of SPS is the fact
that it is almost ideally suited for GPU computing. Here we have used an SPS
algorithm developed specifically in Durham and Geweke (2012) to exploit that
potential.

The multinomial logistic regression model, the focus of this paper, is
important in applied statistics in its own right, and also as a component in
mixture models, Bayesian belief networks, and machine learning. The model
presents a well conditioned likelihood function that renders maximum
likelihood methods straightforward, yet it has been relatively difficult to
attack with posterior simulators -- and hence arguably a bit of an
embarrassment for applied Bayesian statisticians. Recent work by Fr\"{u}%
hwirth and Fr\"{u}hwirth-Schnatter (2007, 2012), Holmes and Held (2006),
Scott (2011) and, especially, Polson et al. (2012) has improved this state
of affairs substantially, using latent variable representations specific to
classes of models that include the multinomial logit.

The SPS algorithm of Durham and Geweke (2012), implemented using Matlab and
a single GPU, led to computation time in the range of 10\% to 100\% of the
computation time required by the algorithm of Polson et al. (2012), which in
turn appears to be the most computationally efficient of alternative
algorithms. Using Matlab and a multicore CPU, the comparison is (very
roughly) reversed, with the algorithm of Polson et al. (2012) having
computation time in the range of 10\% to 100\% of the SPS algorithm But
given the low cost of GPUs -- on the order of US\$250, and the possibility
of having up to 8 GPU's in a single convention desktop machine -- together
with the efficiency of user-friendly software like Matlab, there are no
essential obstacles in moving to GPU computing for applied Bayesian
statisticians. Indeed, some of the case studies are too small to fully
exploit the power of this new platform, as evident in increased efficiency
for SPS/GPU with $10^{5}$ particles as opposed to $10^{4}$ particles.

The SPS algorithm has other attractions that are as significant as
computational efficiency. These advantages are generic, but some are more
specific to the logit model than others.

\begin{enumerate}
\item SPS produces an accurate approximation of the log marginal likelihood
as a by-product. The latent variable algorithms, including all of those just
mentioned, do not. SPS also produces accurate approximations of log
predictive likelihoods, a significant factor in time series models.

\item SPS approximations have a firm foundation in distribution theory. The
algorithm produces a reliable approximation of the standard deviation in the
applicable central limit theorem -- again, as a by-product in the approach
developed in Durham and Geweke (2012). Numerical accuracy in the latent
variable methods for posterior simulation do not do this, and we are not
aware of procedures for ascertaining reliable approximations of accuracy
with these methods that do not entail a significant lengthening of
computation time.

\item More generally, SPS is simple to implement when the likelihood
function can be evaluated in closed form. Indeed, in comparison with
alternatives it can be trivial, and this is the case for the logit model
studied in this paper. By implication, the time from conception to Bayesian
implementation is greatly reduced for this class of models.

\item The ease of implementation, combined with the speed of execution of
the SPS algorithm in a GPU and user-friendly environment, renders the case
for compromises with exact likelihood methods due to exigencies of
application less tenable overall. The same can be said for compromises with
exact subjective Bayesian inference.
\end{enumerate}

\pagebreak

\begin{center}
\textbf{References\medskip }
\end{center}

\begin{description}
\item Albert JH, Chib S (1993). Bayesian analysis of binary and polychotomos
response data. Journal of the American Statistical Association 88: 669-679.

\item Andrieu C, Doucet A, Holenstein A (2010). Particle Markov chain Monte
Carlo. Journal of the Royal Statistical Society, Series B 72: 1--33.

\item Baker JE (1985). Adaptive selection methods for genetic algorithms. In
Grefenstette J (ed.), Proceedings of the International Conference on Genetic
Algorithms and Their Applications, 101--111. Malwah NJ: Erlbaum.

\item Baker JE (1987). Reducing bias and inefficiency in the selection
algorithm. In Grefenstette J (ed.) Genetic Algorithms and Their
Applications, 14--21. New York: Wiley.

\item Chopin N (2002). A sequential particle filter method for static
models. Biometrika 89: 539--551.

\item Chopin N (2004). Central limit theorem for sequential Monte Carlo
methods and its application to Bayesian inference. Annals of Statistics 32:
2385-2411.

\item Chopin N, Jacob PI, Papaspiliopoulis O (2011). SMC$^{2}$: A sequential
Monte Carlo algorithm with particle Markov chain Monte Carlo updates.
Working paper. http://arxiv.org/PS\_cache/arxiv/pdf/1101/1101.1528v2.pdf

\item Del Moral P, Doucet A, Jasra A (2011). On adaptive resampling
strategites for sequential Monte Carlo methods. Bernoulli 18: 252-278.

\item Durham G, Geweke J (2012). Adaptive Sequential Posterior Simulators
for Massively Parallel Computing Environments.
http://www.censoc.uts.edu.au/pdfs/geweke\_ papers/gpu2\_full.pdf.

\item Fahrmeilr L, Tutz G (2001). Multivariate Statistical Modeling based on
Generalized Linear Models. Springer.

\item Foster DP, Stine RA, Waterman RP (1998). Business analysis using
regression. Springer.

\item Flegal JM, Jones GL (2010). Batch means and spectral variance
estimators in markov chain Monte Carlo. Annals of Statistics 38: 1034-1070.

\item Fr\"{u}hwirth-Schnatter S, Fr\"{u}hwirth, R (2007). Auxiliary mixture
sampling with applications to logistic models. Computational Statistics and
Data Analysis 51: 3509 -- 3582.

\item Fr\"{u}hwirth-Schnatter S, Fr\"{u}hwirth, R (2012). Bayesian inference
in the multinomial logit model. Austrian Journal of Statistics 41: 27-43.

\item Geweke J, Keane M (2007). Smothly mixing regressions. Journal of
Econometrics 138: 252-290.

\item Geweke J, Keane M, Runkle D (1994). Alternative computational
approachesto inference in the multinomial probit mdoel. Review of Economics
and Statistics 76: 609-632.

\item Gordon NJ, Salmond DJ, Smith AFM (1993). Novel approach to nonlinear /
non-Gaussian Bayesian state estimation. IEEE Proceedings F on Radar and
Signal Processing 140 (2): 107-113.

\item Gramacy RB, Polson NG (2012). Simulation-based regularized logistic
regression. Bayesian Analysis 7: 567-589.

\item Green WH (2003). Econometric Analysis. Fifth Edition. Prentice-Hall.

\item Herbst E, Schorfheide F (2012). Sequential Monte Carlo sampling for
DSGE models. http://www.ssc.upenn.edu/\symbol{126}schorf/papers/smc%
\_paper.pdf

\item Holmes C, Held L (2006). Bayesian auxiliary variable models for binary
and multinomial regression. Bayesian Analysis 1: 145-168.

\item Jacobs RA, Jordan MI, Nowlan SJ (1991). Adaptive mixtures of local
experts. Neural Computation 3: 79-87.

\item Jiang WX, Tanner MA (1999).Hierarchical mixtures-of-experts for
exponential family regression models: Approximation and maximum likelihood
estimation. Annals of Statistics 27: 987-1011.

\item Kong A, Liu JS, Wong WH (1994). Sequential imputations and Bayesian
missing data problems. Journal of the American Statistical Association
89:278--288.

\item Liu JS, Chen R (1995). Blind deconvolution via sequential imputations.
Journal of the American Statistical Association 90: 567--576.

\item Liu JS, Chen R (1998). Sequential Monte Carlo methods for dynamic
systems. Journal of the American Statistical Association 93: 1032--1044.

\item Polson NG, Scot JG, Windle J (2012). Bayesian inference for logistic
models using Polya-Gamma latent variables.
http://faculty.chicagobooth.edu/nicholas.polson /research/papers/polyaG.pdf

\item Scott SL (2011). Data augmentation, frequentist estimation, and the
Bayesian analysis of multinomial logit models. Statistical Papers 52; 87-109.

\item Zellner A (1986). On assessing prior distributions and Bayesian
regression analysis with g-prior distributions. In: Goel P, Zellner A
(eds.), Bayesian Inference and Decision Techniques: Studies in Bayesian
Econometrics and Statististics. 6 233--243. North-Holland.
\end{description}

\end{document}